\documentclass[conference]{IEEEtran}
\IEEEoverridecommandlockouts
\usepackage{cite,url}
\usepackage{amsmath,amssymb,amsfonts}
\usepackage{algorithmic}
\usepackage{graphicx}
\usepackage{textcomp}
\usepackage{xcolor}

\def\BibTeX{{\rm B\kern-.05em{\sc i\kern-.025em b}\kern-.08em
    T\kern-.1667em\lower.7ex\hbox{E}\kern-.125emX}}
\begin{document}
\title{PrimaDNN':A Characteristics-aware DNN Customization for Singing Technique Detection}
\def\firstauthor{Yuya Yamamoto}
\def\secondauthor{Juhan Nam}
\def\thirdauthor{Hiroko Terasawa}

\author{\IEEEauthorblockN{Yuya Yamamoto}
\IEEEauthorblockA{\textit{Doctoral Program in Informatics} \\
\textit{University of Tsukuba,}\\
Ibaraki, Japan \\
s2130507@s.tsukuba.ac.jp}
\and
\IEEEauthorblockN{Juhan Nam}
\IEEEauthorblockA{\textit{Graduate School of Culture Technology} \\
\textit{KAIST}\\
Daejeon, South Korea \\
juhan.nam@kaist.ac.kr,}
\and
\IEEEauthorblockN{Hiroko Terasawa}
\IEEEauthorblockA{\textit{Doctoral Program in Informatics} \\
\textit{University of Tsukuba,}\\
Ibaraki, Japan \\
terasawa@slis.tsukuba.ac.jp}
}

\newcommand\figref{Fig.~\ref}
\newcommand\tabref{Table.~\ref}
\newcommand\chatgpt{\textcolor{red}}
\maketitle

\begin{abstract}

Professional vocalists modulate their voice timbre or pitch to make their vocal performance more expressive. Such fluctuations are called singing techniques. Automatic detection of singing techniques from audio tracks can be beneficial to understand how each singer expresses the performance, yet it can also be difficult due to the wide variety of the singing techniques. A deep neural network (DNN) model can handle such variety; however, there might be a possibility that considering the characteristics of the data improves the performance of singing technique detection. In this paper, we propose PrimaDNN, a CRNN model with a characteristics-oriented improvement. The features of the model are: 1) input feature representation based on auxiliary pitch information and multi-resolution mel spectrograms, 2) Convolution module based on the Squeeze-and-excitation (SENet) and the Instance normalization.
In the results of J-POP singing technique detection, PrimaDNN achieved the best results of 44.9\% at the overall macro-F measure, compared to conventional works.
We also found that the contribution of each component varies depending on the type of singing technique. 
\end{abstract}


\section{Introduction}

\begin{figure}
    \centering
    \includegraphics[width=\columnwidth]{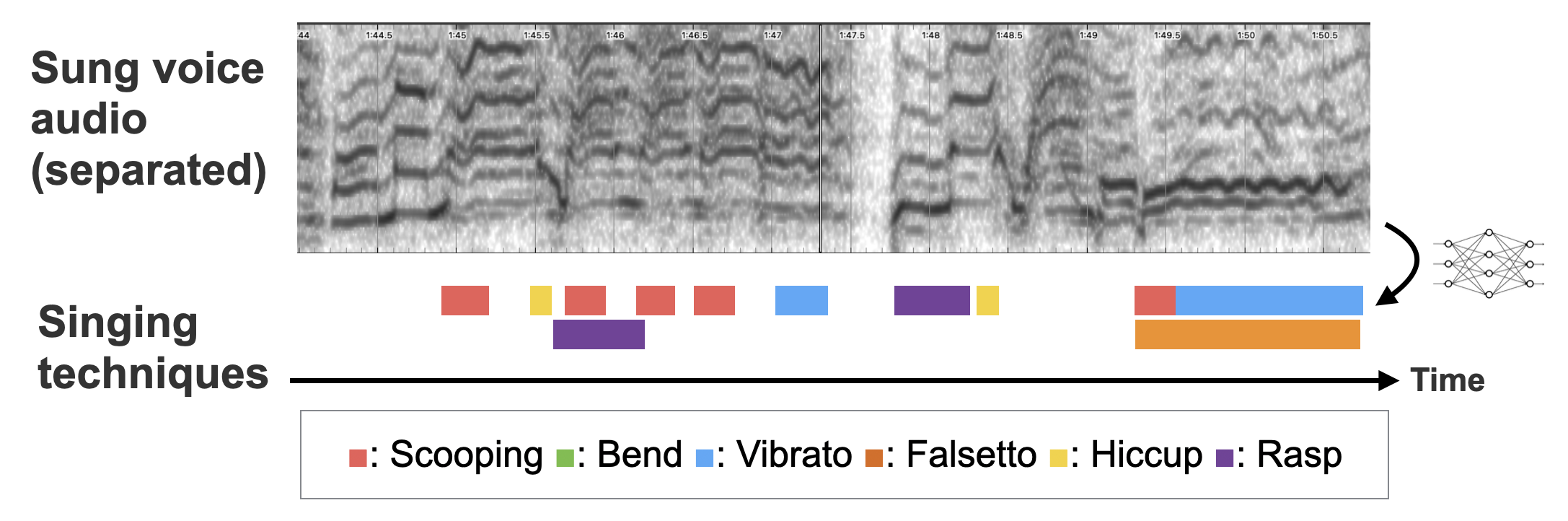}
    \caption{The overview of singing technique detection. It is a multi-label and frame-wise classification that locates and identifies the singing techniques given a sung audio clip.}
    \label{fig:task}
\end{figure}

A singing voice is one of the most essential elements of music, providing impactful emotional expressions through melody and lyrics. In particular, in popular music, the role of the singing voice is even more critical, as the vocal quality and unique style of singers are crucial in attracting people. Vocals mainly consist of singer individuality (i.e., vocal fold vibration and vocal tract resonance) and singing expressions (i.e., fine control of pitch, timbre, and loudness). We define the latter as singing techniques, which are the singing voice versions of extended playing techniques.
Automatic identification of singing techniques from sung voice tracks can contribute to the understanding of different singing styles, and can have applications in music discovery, vocal training, and user-generated content. 
It can also help to reduce the laborious and specialized process of analyzing singing techniques for billions of songs.
In our previous work \cite{yamamoto2022analysis}, we tackled singing technique detection from real-world repertoires in J-POP, where the demand for singing technique analysis is high. 
Figure \ref{fig:task} shows a quick overview of singing technique detection. It is a multi-class, multi-label classification in each analysis frame, where the input is an audio track of a singing voice, and the output is a timeline of the appearance of singing techniques. The difficulties of the task lie in localization and classification, where a wide variety of noise and fluctuation exists.

\begin{figure}
    \centering
    \includegraphics[width=\columnwidth]{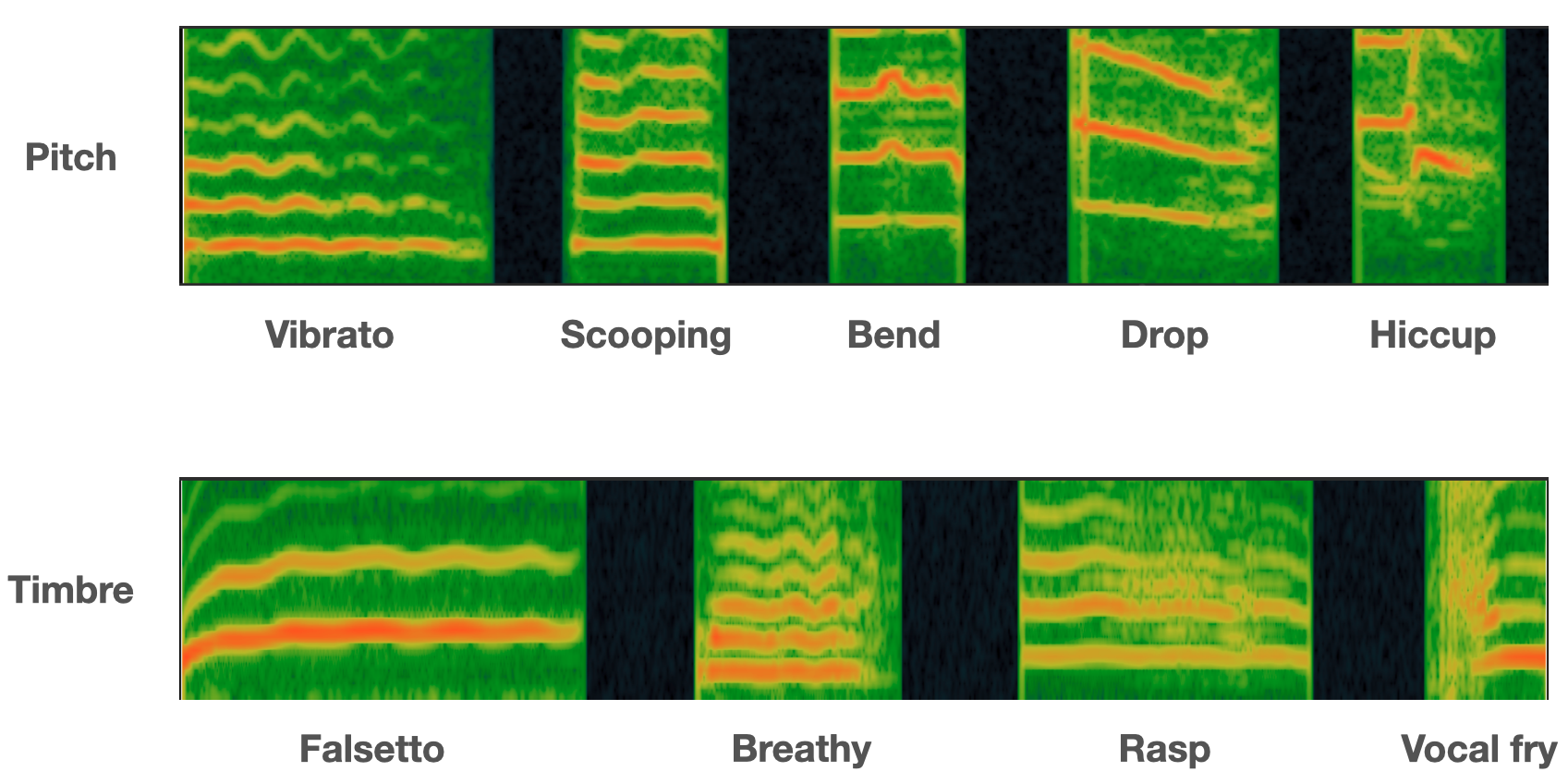}
    \caption{The spectrograms of nine singing techniques that we treat in the experiment. The upper and the lower show pitchy and timbral techniques, respectively.}
    \label{fig:spec_technique}
\end{figure}

Recently, deep neural networks have achieved high performance on identification tasks, even in challenging conditions. Therefore, in our previous work \cite{yamamoto2022analysis}, we adopted the CRNN model, which is one of the succeeding DNN architectures in many audio and music identification tasks \cite{adavanne2018sound,nishikimi2021audio,choi2017convolutional} in a series of experiments. We found that DNN models considering the characteristics of data have the potential to improve identification results. Figure \ref{fig:spec_technique} shows the spectrogram of singing techniques that we analyzed in this paper.
Each technique displays different pitch modulation or spectral patterns. For instance, Vibrato' shows a sinusoidal-shaped periodic pitch modulation, whereas Scooping' shows an S-shaped continuous pitch change. In terms of timbral techniques, Vocal fry' exhibits fast pulsive patterns, while Rasp' shows sub-harmonics. Therefore, the model must capture such a wide variety of acoustic characteristics to improve the detection performance.

In this paper, we reconsider the architecture of DNN by considering the characteristics of singing techniques. To achieve the aforementioned improvement, our model focuses on the following two aspects: 1) feature representation that captures the wide variety and 2) the mechanism that suppresses the effect of features that have nothing to do with the desired targets. For the first aspect, we adopt two approaches: multi-resolution mel spectrograms to capture various types of modulation, and mel-band pitchgram that explicitly informs the sung pitch heights. For the second aspect, we also adopt two approaches: Squeeze-and-Excitation network to dynamically select the important feature map on the convolutional layer, and Instance normalization to prevent instance-specific mean and co-variance shift that may impede the capturing of target features. We named the DNN model PrimaDNN' (pronounced prima-don-na)\footnote{We provide the detail at \url{https://yamathcy.github.io/eusipco23primadnn/}.}.

\section{Architecture}
\begin{figure*}
    \centering
    \includegraphics[width=\textwidth]{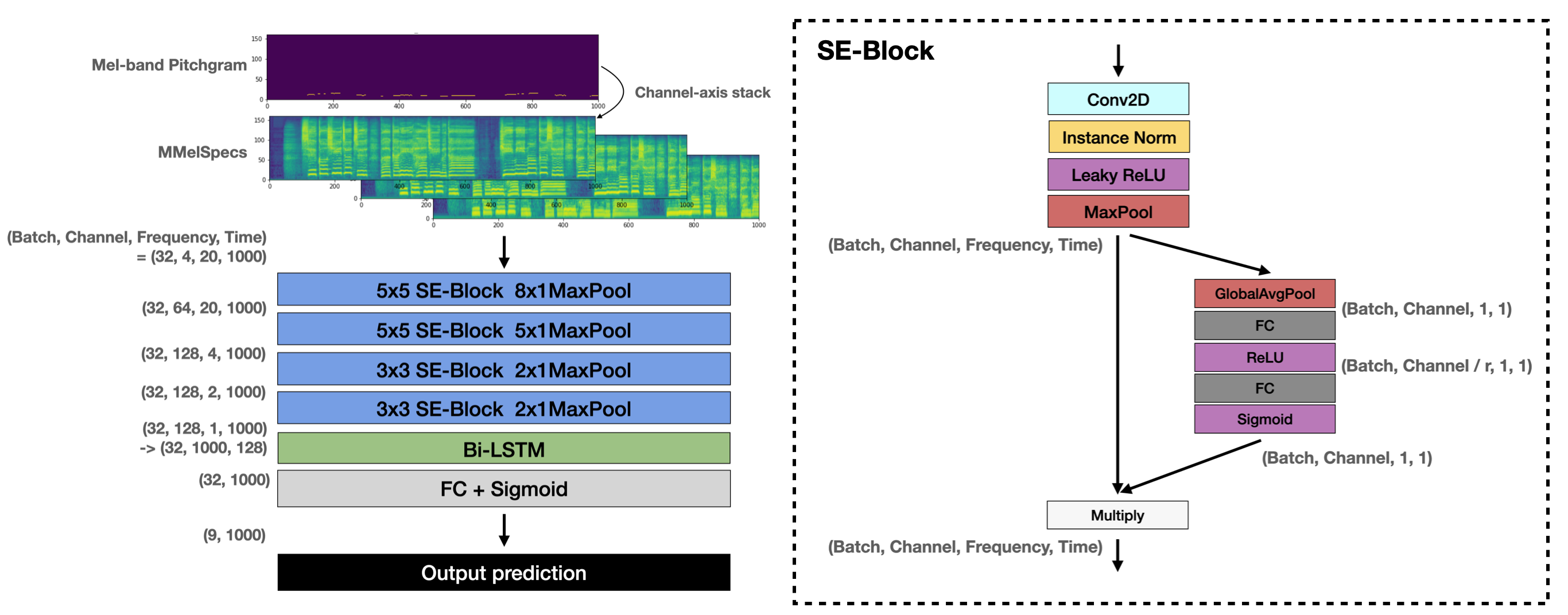}
    \caption{The overview of PrimaDNN' model. (Left) the diagrams of architecture. (Right) the diagrams of SE-Block.}
    \label{fig:overview}
\end{figure*}

\figref{fig:overview} shows our proposed PrimaDNN' model.  
It is following a CRNN model that has four convolutional layers, 1 Bi-directional LSTM layer, 1 Fully-connected (FC) layer and 1 sigmoid activation layer. 
Only in the inference time, the output is binarized by thresholding with the value of 0.5.

\subsection{Input feature}
To overcome issues we use stacked multi-resolution mel spectrograms and 2D Mel band pitchgram for the input feature.

\textbf{Multi-resolution mel spectrograms (MMelSpecs)} are made by stacking three mel spectrograms which have different time-frequency resolutions with each other, in order to adapt wide modulation patterns both on time and frequency bands of singing techniques\cite{yamamoto2021investigating}.
We adopt window sizes of (2048, 1024, and 512) for short-time Fourier Transform (STFT) with Hann-window, maintaining the same size for all mel spectrograms by zero-padding and applying fixed hop size.
All of these mel spectrograms have a frequency dimension of 160 and each frame length of 10 ms.

In addition, we stacked \textbf{Mel-band pitchgram}\cite{hsieh2020addressing, yamamoto2022analysis} on MMelSpecs. 
It has the same frequency dimensions as the input mel spectrograms and has one-hot where the pitch frequency exists. 
We use the pitch that is automatically estimated by CREPE\cite{kim2018crepe}, one of the state-of-the-art pitch extraction. 
Note that although the conventional work\cite{yamamoto2022analysis} that using ground-truth pitch shows the best performance, using CREPE also shows competitive results.

\subsection{DNN architecture}
We adopt \textbf{Squeeze-and-Excitation (SENet)}\cite{hu2018squeeze} and Instance normalization\cite{ulyanov2016instance} for customization of the convolution layers of CRNN model.
SENet is originally proposed in image domain, in order to enhance the representative power of a neural network by feature re-calibration that emphasizes informative features and suppresses useless ones. 
As the right side of \figref{fig:overview} shows, SENet squeezes the input feature maps by Global average pooling, then reduces the channel dimension with a ratio of $r$ on the first fully connected (FC) layer.
Finally, the second FC layer rescales the channel dimension and outputs the importance of each feature map, which has a value range of $[ 0,1 ]$.
In all of the conditions that use SE, we empirically set $r$ to $2$ from the grid search on the range of $[16,8,4,2]$.

For the normalization method, we use \textbf{instance normalization (IN)} instead of batch normalization (BN) everywhere in the network with the purpose of leading the model to focus on features relates to singing techniques.
IN prevents instance-specific mean and covariance shift simplifying the learning process. 
IN is mainly used in style transfer to disentangle the content and style\cite{pan2018two}. In the audio domain, it is used for speaker emotion recognition\cite{zhang2021autoencoder}, speaker conversion \cite{Chou2019} to suppress the effect of non-target attributes (e.g., speaker information, speech content, etc.) 
We expect that IN can get invariance of irrelevant attributes to singing techniques (e.g., singer identity, vocal mixing style, quality of vocal separation, vocal note density, etc.)

We trained the model using \textbf{Focal loss} \cite{lin2017focal}. Singing technique detection is difficult due to data imbalance, which can negatively affect detection performance. Focal loss addresses this by focusing training on hard examples (i.e., the frames where singing techniques appear in this case) and down-weighting the loss assigned to easy examples.
The equation of Focal loss given the output activation $p$, is as follows:
\begin{equation}
    L_{fl}(p_t) = -\alpha (1-p_t)^\gamma \log (p_t)\\
\end{equation}
\begin{equation}
    p_t = \left\{
\begin{array}{ll}
 p & label = 1 \\
1-p & otherwise
\end{array}
\right.
\end{equation}
 $\alpha \in [0, 1] $ is a weighting factor for balancing the importance of positive and negative examples, and the term $(1 - p_t)^{\gamma}$ is a modulating factor, with $\gamma$ controlling the rate
of dominant examples. 
We conducted a grid search on the range of $\alpha = [0.1, 0.13, 0.15, 0.2, 0.25]$ and $\gamma = [1, 1.33, 1.66, 2.0]$ and set $\alpha$ to 0.13 and $\gamma$ to 1.33 for all conditions in the work that used focal loss.


\section{Experiments}
We conduct an experiment of nine-way singing technique detection.
\subsection{dataset}
We evaluated the proposed architecture using the COSIAN dataset \cite{yamamoto2022analysis}, which includes 168 tracks of four famous hit songs sung by 42 solo singers of both genders. For the experiment, we selected the most common nine techniques. We processed the vocal tracks by sampling them to 44.1 kHz, separating them using Demucs v3 \cite{defossez2021hybrid}, and segmenting them into 10-second non-overlapping parts. We then obtained input features using MMelSpecs by applying short-time Fourier transform (STFT) with a 2048-sample Hann window and a hop size of 10 ms.

\subsection{evaluation}
To evaluate the performance of our proposed architecture, we conducted singer-wise seven-fold cross-validation, as in our previous work \cite{yamamoto2022analysis}. We divided the singers into seven groups and organized the dataset into training, validation, and test sets with a ratio of 5:1:1 for each set.

To account for label imbalances between singers, we used the nine most common singing techniques ('bend', 'breathy', 'drop', 'falsetto', 'hiccup', 'rasp', 'scooping', 'vibrato', and 'vocal fry'), which appeared in every fold of the cross-validation.

Our evaluation metrics included segment-based recall (\textbf{R}), precision (\textbf{P}), macro-F-measure (\textbf{Macro-F}), and micro-F-measure (\textbf{Micro-F}) \cite{mesaros2016metrics}, as well as the F-measure for each singing technique. We calculated these metrics using sed\_eval\footnote{\url{https://tut-arg.github.io/sed_eval/index.html}}. The macro-F-measure represents the class-wise average of the F-measure, while the micro-F-measure represents the instance-wise average. We set the segment length to 100 ms for our evaluation.

\section{Results and discussion}

\subsection{Comparison with baseline}
First, we compare our proposed model with baseline models.
As baselines, we prepared four conventional models. 1) \textit{eGeMaps LSTM}\cite{atmaja2020differences}: eGeMaps \cite{eyben2015geneva} is a feature set used in speech emotion recognition tasks. It consists of 25 low-level descriptors for each frame. In this model, we used eGeMaps as an input feature and fed it to an LSTM model. 2) \textit{CRNN} \cite{imoto2021impact} A simple CRNN model whose input is a 64-dimensional Mel spectrogram and has three convolutional layers, one Bi-GRU layer and one FC layer. 3) \textit{CNN Self-Attention} \cite{imoto2021impact, kong2020sound} Instead of Bi-GRU layer, multi-head attention is applied. This model achieved the best performance on sound event detection with data imbalance situation\cite{imoto2021impact}. 
In addition, we also compared with \textit{CRNN+PitchFocal}, a CRNN that is fed the Mel-band pitchgram and applied Focal loss, both of which improved the performance of singing technique detection\cite{yamamoto2022analysis}.
All models were trained using the RAdam optimizer \cite{liuvariance2019radam} with a learning rate of 1e-3. Training stopped if the value of the loss function on the validation set did not improve for 20 epochs.

We used binary cross entropy (BCE) as the loss function for \textit{eGeMaPS}, \textit{CRNN}, \textit{CNN Self-Attention} and Focal loss\cite{lin2017focal} for \textit{CRNN+PitchFocal} as in the original work. 

\tabref{tab:exp_results} displays the results of the experiment. 
PrimaDNN' achieved 44.9\% at \textit{Macro-F}, 60.6 \% at \textit{Micro-F}, 43.8\% at \textit{Precision} and 48.3\% at \textit{Recall}, respectively, as shown in the bottom of the table.
These results indicate that PrimaDNN' outperformed the conventional models in all of the metrics.
 \begin{table}[!t] 
\centering
\caption{The results of singing technique detection.}
 \label{tab:exp_results}
\begin{tabular}{ccccc}
\hline
 & \textbf{Macro-F} & \textbf{Micro-F}  & \textbf{P} & \textbf{R}   \\ \hline 
eGeMaps LSTM & 9.2\% & 6.3\% &  11.3\% & 1.6\%    \\ 
CRNN & 37.7\%  & 56.3\% & 42.2\% & 39.2\% \\ 
CRNN+PitchFocal  & 40.2\% & 55.1\% & 37.7\% & 48.0\%          \\ 
CNN Self-Attention& 42.0\%& 59.3\% & 43.4\% & 47.7\%   \\ 
\hline
PrimaDNN' (ours) & \textbf{44.9}\% &\textbf{ 60.6}\% &  \textbf{43.8}\% & \textbf{48.3}\%    \\ 
\end{tabular}
\end{table}
\subsection{Ablation study}
In order to understand the contribution of each component in our model, we conducted an ablation study by comparing our full model with several modified versions, as outlined below:
\begin{itemize}
    \item \textbf{Single resolution}: Uses only a single resolution mel spectrogram that was processed by STFT with window length of 2048.
    \item \textbf{No SE}: Remove the SE blocks from each convolution layer.
    \item \textbf{BN}: Replace IN with Batch Normalization (BN).
        \item \textbf{No pitch}: Removes mel band pitchgram from input.
    \item \textbf{3x3}: Adopt 3x3 for the kernel size of all convolution layer. (i.e., instead of 5x5 for the first and the second convolution layer.)
\end{itemize}
The experiments showed that the \textit{full} model outperformed all the modified versions in terms of both \textit{Macro-F} and \textit{Micro-F}. 

\begin{figure}[!ht]
    \centering
    \includegraphics[width=\columnwidth]{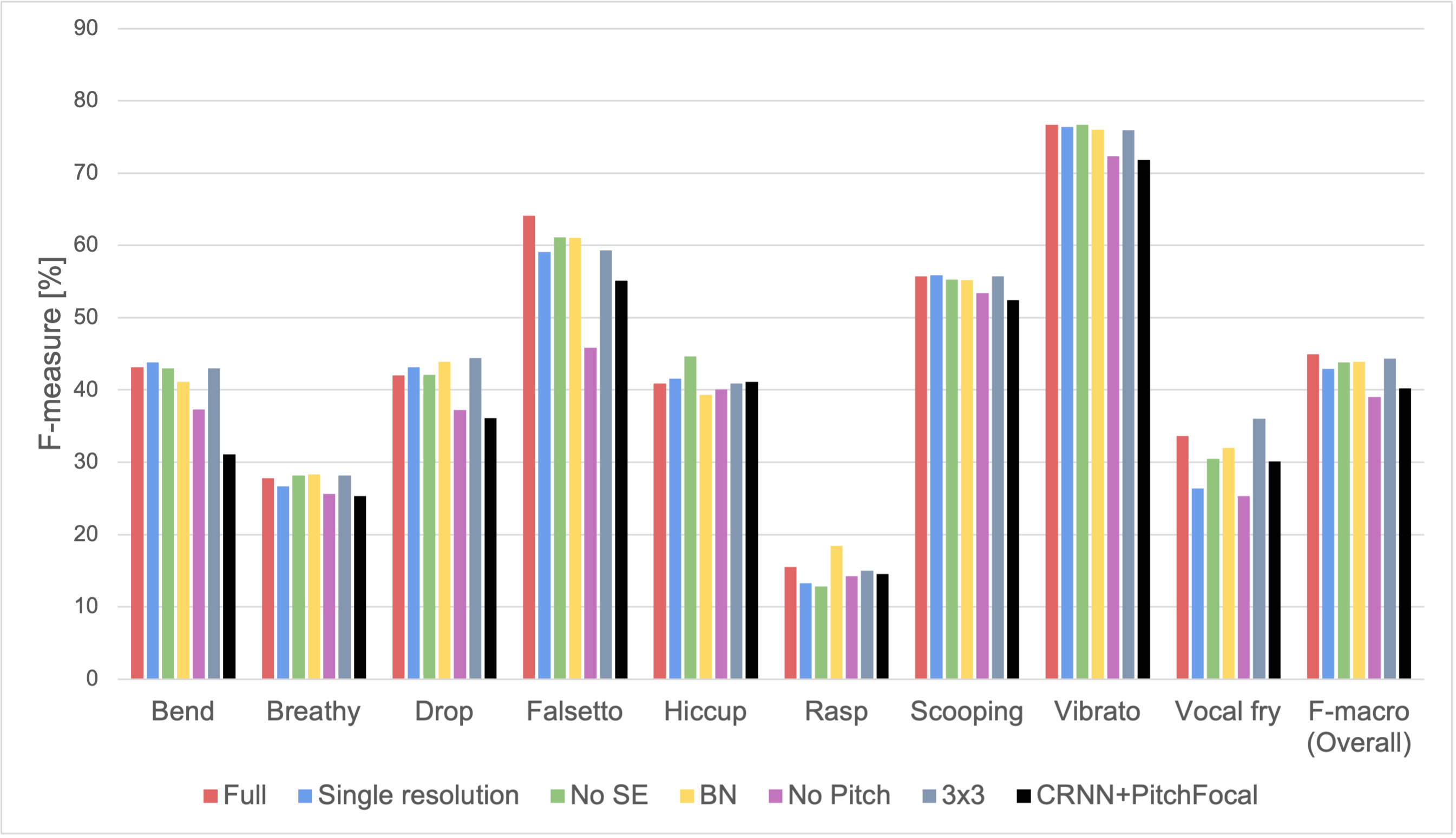}
    \caption{The technique-wise F-measures for each method in ablation study.}
    \label{fig:technique_wise}
\end{figure}

We further examine the class-wise F-measure and compare it with our previous best model (CRNN-PitchFocal) \cite{yamamoto2022analysis}. As shown in \figref{fig:technique_wise}, our model outperforms the previous one in most techniquesd.    
The main difference between our model and the previous one is the frequency dimension of the input feature, where we adopted a higher resolution of 160. 
This improvement led to better performance in detecting pitchy techniques such as vibrato', bend', drop', and scooping', indicating that higher frequency resolution better represents fine pitch fluctuation.

We also found that the multi-resolution spectrogram improved the detection of vocal fry' compared to using a single resolution (i.e., with a window length of 2048 only). Vocal fry' has a pulsive modulation pattern as shown at the bottom of \figref{fig:spec_technique}. Combining spectrograms with fine temporal resolution helps capture its characteristics. Additionally, instance normalization helped with the detection of `falsetto'.

The \textit{3x3} condition performed similarly to the \textit{full} model. However, it showed better performance on techniques with shorter duration (e.g., drop' and vocal fry'), but worse performance on techniques with longer duration (e.g., falsetto', rasp', and `vibrato'), compared to the \textit{full} model. This indicates that the size of the receptive fields affects the detection performance of different techniques.

 \begin{table}[!t]
\centering
\label{tab:ablation}
\caption{The results of singing technique detection.}
\begin{tabular}{ccccc}
\hline
 & \textbf{Macro-F} & \textbf{Micro-F}  & \textbf{P} & \textbf{R}   \\ \hline 
PrimaDNN'(Full) & \textbf{44.9\%} &\textbf{ 60.6\%} &  43.8\% & 48.3\%    \\ 
\hline
No pitch & 39.0\%  & 54.8\% & 36.6\% & 47.3\% \\ 
Single resolution & 42.9\% & 60.2\% & 44.1\% & 46.6\% \\ 
No SE & 43.8\%& 60.3\% & 43.0\% & 48.1\%  \\ 
BN   & 43.9\% & 59.6\% & \textbf{44.6\%} & 48.1\% \\ 
3x3 & 44.3\% & 60.0\% & 43.2\% & \textbf{48.8\%} \\ \hline
\end{tabular}
\end{table}


\subsection{Detection examples}

\begin{figure*}[!ht]
    \centering
    \includegraphics[width=\textwidth]{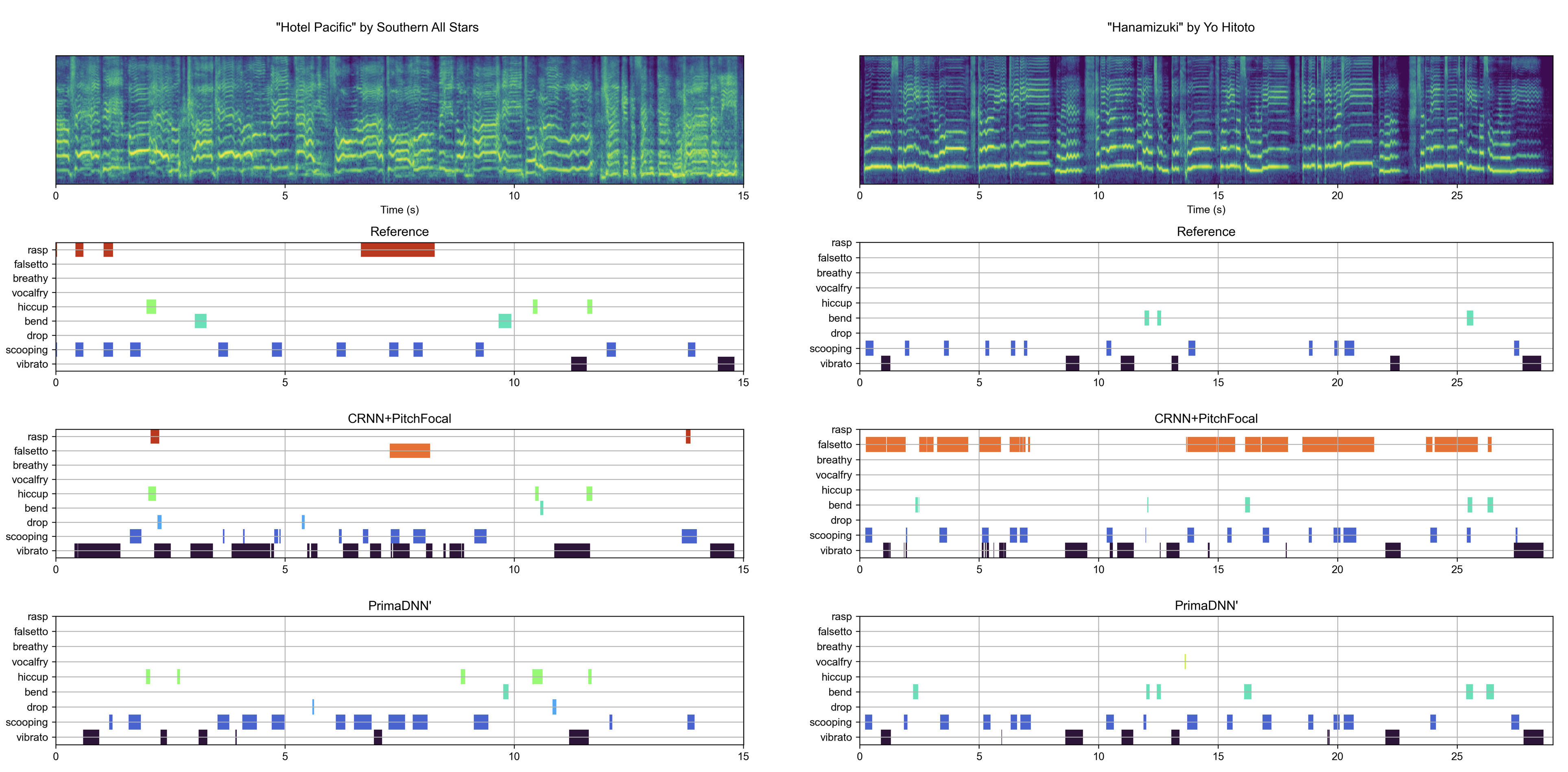}
    \caption{Detection examples. From above, spectrogram, reference annotation, estimation of CRNN+PitchFocal\cite{yamamoto2022analysis}, and estimation of PrimaDNN' in each row. (Left) Comparison on "Hotel Pacific" by Southern All Stars. (Right) Comparison on "Hanamizuki" by Yo Hitoto.}
    \label{fig:ex}
\end{figure*}

In order to investigate the detailed detection performance, we present examples of detections made by CRNN+PitchFocal and PrimaDNN' with reference annotations in Figure \ref{fig:ex}. 
The example on the left side of the figure depicts a song with many fine fluctuations and note changes. CRNN+PitchFocal detected many false positives in the vibrato' category at the positions of note transition, whereas PrimaDNN' was able to suppress such false positives.
The example on the right side of the figure depicts a song with a slow tempo and mellow mood sung by a female singer. 
As the figure shows, the section displayed does not have any falsetto', but CRNN+PitchFocal detected them as false positives. 
In contrast, PrimaDNN' did not detect any `falsetto' sections as per the reference annotations, indicating that it may be more powerful and robust than CRNN+PitchFocal.

\section{Conclusion}
This paper introduces Prima-DNN', a DNN architecture that takes into account the specific characteristics of singing techniques. It employs multi-resolution mel spectrograms and Mel-band pitchgram for input features, Squeeze-and-Excitation network, and Instance normalization for convolutional layers. The proposed model achieves the best performance on the nine-way detection of singing techniques on the COSIAN dataset. Furthermore, it demonstrates an ability to reduce false negatives for difficult patterns such as those between fast passages and vibrato and non-falsetto singing at high pitch notes and falsetto.

The study \cite{yamamoto2021analysis} suggests that there are certain correlations between the appearance of singing techniques and musical context (e.g., note pitch and duration, phoneme of lyrics, the position of phrase, singer, etc.). Therefore, for future works, it is proposed to combine features related to other musical components such as musical notes, lyrics, and singer information. This could be done through the use of pre-trained features (e.g., Wav2Vec2.0\cite{baevski2020wav2vec}, ECAPA-TDNN speaker embedding\cite{desplanques2020ecapa}) or multi-task learning.
\bibliography{r}
\bibliographystyle{IEEEtran}

\end{document}